\documentstyle[12pt,epsfig]{article}

\def\be{\begin{equation}}
\def\ee{\end{equation}}

\begin{document}

\begin{titlepage}
\setlength{\textwidth}{5.0in}
\setlength{\textheight}{7.5in}
\setlength{\parskip}{0.0in}
\setlength{\baselineskip}{18.2pt}
\setlength{\footskip}{0.5in}
\setlength{\footheight}{0in}

\renewcommand{\thefootnote}{\fnsymbol{footnote}}


\vspace{0.3cm}

\begin{center}
{\Large\bf The geodesic motion near hypersurfaces in the warped products spacetime}
\end{center}

\begin{center}
Jaedong Choi\footnote{Electronic address: choijdong@gmail.com}$^{1}$,
Yong-Wan Kim\footnote{Electronic address: ywkim65@gmail.com}$^{2}$
and Young-Jai Park$\footnote{Electronic address: yjpark@sogang.ac.kr}^{2,3}$\par
\end{center}

\begin{center}
{${}^{1}$Department of Basic Science, Korea Air Force Academy 363-849, Korea}
\par {${}^{2}$Center for Quantum Spacetime, Sogang
University, Seoul 121-742, Korea}\par {${}^{3}$Department of
Physics, Sogang University, Seoul 121-742, Korea}
\end{center}

\vskip 0.5cm
\begin{center}
{\today}
\end{center}

\vfill

\begin{abstract}
In the framework of Lorentzian multiply warped products we study
the Gibbons-Maeda-Garfinkle-Horowitz-Strominger (GMGHS) spacetime
near hypersurfaces in the interior of the event horizon. We also
investigate the geodesic motion in hypersurfaces.
\end{abstract}

\vskip20pt

PACS numbers: 04.70.Bw, 04.50.Kd
\vskip15pt
Keywords: Black hole interior, Multiply warped products \\
\end{titlepage}

\newpage

\section{ Introduction}
\setcounter{equation}{0}
\renewcommand{\theequation}{\arabic{section}.\arabic{equation}}
The concept of a warped product manifold was introduced to provide a
class of complete Riemannian manifolds with negative curvature
everywhere~\cite{bo}, and was developed to point out that several of
the well-known exact solutions to Einstein field equations are
pseudo-Riemannian warped products~\cite{be}. Furthermore, certain
causal and completeness properties of a spacetime can be determined
by the presence of a warped product structure~\cite{bpe}, and a
general theory Lorentzian multiply warped were applied to discuss
the Schwarzschild spacetime in the interior of the event horizon
~\cite{choi00,choi01,choi02,choi03}. The
role of warped products in the study of exact solutions to
Einstein's equations are now firmly established to generate interest
in other areas of geometry. \vskip10pt

On the other hand, there were enormous interests in the spherically
symmetric static charged black holes in the four-dimensional
heterotic string theory, which have similarities as well as
differences with the Reissner-Nordstr\"om black hole in general
relativity~\cite{Horowitz1992}. By turning antisymmetric tensor
gauge fields off, the static charged black hole solution was found
by Gibbons, Maeda~\cite{gm}, and by Garfinkle, Horowitz,
Strominger~\cite{ghs}, independently. Recently, null geodesics and
hidden symmetries in the Sen black hole was investigated by Hioki
and Miyamoto~\cite{hm08}, which is reduced to the GMGHS black hole
in the nonrotating limit. Gad~\cite{gad} also studied geodesic and
geodesic deviation of the GMGHS black hole solution. Very recently,
Fernando~\cite{Fernando} fully investigated null geodesic motions of
the same solution both in the Einstein and string frame. However,
the studies of these null geodesics solutions are mainly based on
the exterior region of the event horizon. In a Lorentzian multiply
warped product spacetime, by exchanging timelike and spacelike
coordinates, we are interest in the interior region of the event
horizon. \vskip10pt

In this paper we study the GMGHS interior spacetime of the framework of
Lorentzian multiply warped products. We also investigate the
geodesic motion near hypersurfaces of this spacetime. We
shall use geometrized units, i.e., G = c = 1, for notational
convenience.


\section{GMGHS black hole in the framework of warped products}
\setcounter{equation}{0}
\renewcommand{\theequation}{\arabic{section}.\arabic{equation}}

 The four dimensional low energy GMGHS action obtained from
heterotic string theory~\cite{Horowitz1992,gm,ghs} is given by
 \be
 S=\frac{1}{16\pi}\int d^4x
 \sqrt{-g}\Big[R-2(\nabla\Phi)^2-e^{-2\Phi}F_{\mu\nu}F^{\mu\nu}\Big],
 \ee
where $\Phi$ is a dilaton field and $F_{\mu\nu}$ the Maxwell field
strength tensor. The low energy limit of string theory includes a
scalar dilaton field, which is massless in all finite orders of
perturbation theory~\cite{GSW}. In particular, the dilaton is
coupled with the Maxwell field and it affects the geometry of
spacetime, making different from the Reissner-Nordstr\"om solution
of the Einstein-Maxwell theory.\par

The GMGHS solution of the Einstein field equations in Einstein frame
describes the geometry exterior to a spherically symmetric static
charged black hole as
\begin{equation}\label{rs}
ds^2=-\Bigl({1-\frac{2m}{r}}\Bigr)dt^2+\Bigl({1-\frac{2m}{r}}\Bigr)^{-1}dr^2
     +r^2\Bigl(1-{\frac{\alpha}{r}}\Bigr)(d\theta^2+\sin^2\theta d\phi^2),
\end{equation}
where
\begin{equation}
\label{r00w}e^{2\Phi}=1-\frac{\alpha}{r}, ~~F_{rt}=\frac{Q}{r^2}
\end{equation}
with $\alpha=Q^2/m$.  The parameters $m$ and $Q$ are mass and
charge respectively. Note that the metric in the $t$-$r$ plane is
identical to the Schwarzschild case. As like in the Schwarzschild
spacetime, the GMGHS has an event horizon at $r=2m$. We also note
that the area of the sphere of the GMGHS black hole, defined by
$\int d\theta d\phi \sqrt{g_{\theta\theta}g_{\phi\phi}}$, is
smaller than the Schwarzschild spacetime by an amount depending on
the charge. In particular, the area of the sphere approaches zero
as $r\rightarrow\alpha$, leading to a surface singularity as
\begin{equation}
R=\frac{\alpha^2(r-2m)}{2r^3(r-\alpha)^2}.
\end{equation}
As far as the case of $\alpha\le 2m$, the singular surface remains
inside the event horizon so that the Penrose diagram is identical
to the Schwarzschild spacetime. Also the case of  $\alpha=2m$,
which implies $Q^2=2m^2$ is the extremal limit which the event
horizon and the surface singularity meet.

 It is also interesting to note that in the GMGHS solution
(\ref{rs}), the limit of $Q\rightarrow 0$ makes the dilaton
vanishing so that the only black hole solution is the
Schwarzschild one. On the contrary, every solution with nonzero
Maxwell fields must have a nonconstant dilaton which will alter
the geometry~\cite{Horowitz1992,gm,ghs}. This is the main
difference from the Reissner-Nordstr\"om solution in which case
can be embedded into the 5 dimensional 3-parameter class of
charged solitons {\it a la} Kaluza-Klein~\cite{Nodvik,liu1,liu2}.
\vskip10pt

On the other hand, the line element for the GMGHS metric for the
interior region $r<2m$ can be described by
\begin{equation}\label{inside}
ds^2=-\Bigl({\frac{2m}{r}}-1\Bigr)^{-1}dr^2+\Bigl({\frac{2m}{r}-1}\Bigr)dt^2
     +r^2\Bigl(1-{\frac{\alpha}{r}}\Bigr)(d\theta^2+\sin^2\theta d\phi^2),
\end{equation}
where $r$ and $t$ are now new temporal and spacial variables,
respectively. A multiply warped product manifold, denoted by
$M=(B\times F_1\times...\times F_{n}, g)$, consists of the
Riemannian base manifold $(B, g_B)$ and fibers $(F_i,g_i)$
($i=1,...,n$) associated with the Lorentzian metric~\cite{choi00}.
In particular, for the specific case of $(B=R,~g_B=-d\mu^{2})$,
the GMGHS metric (\ref{inside}) can be rewritten as a multiply
warped products $(a, b)\times_{f_1}R\times_{f_2} S^2$ by making
use of a lapse function
\begin{equation}
N^{2}=\frac{r_{H}-r}{r}
\label{schlapse2}
\end{equation}
as well as warping functions given by $f_1$ and $f_2$ as follows
\begin{eqnarray}\label{fs}
f_{1}(\mu)&=&\left(\frac{2m}{F^{-1}(\mu)}-1\right)^{1/2},\nonumber\\
f_{2}(\mu)&=&\left(F^{-1}(\mu)^2-\alpha F^{-1}(\mu)\right)^{1/2}.
\label{schf1f2}
\end{eqnarray}
The lapse function (\ref{schlapse2}) is well
defined in the region $r<r_{H}(=2m)$ to rewrite it as a multiply
warped products spacetime by defining a new coordinate $\mu$ as
follows
\begin{equation}
\mu=\int_{0}^{r}\frac{dx~x^{1/2}}{(r_{H}-x)^{1/2}}=F(r).
\label{musch}
\end{equation}
Setting the integration constant zero as $r\rightarrow 0$, we have
\begin{equation}
\mu=2m\cos^{-1}\left(\frac{r_{H}-r}{r_{H}}\right)-[(r_{H}-r)r]^{1/2},
\label{schsolmu}
\end{equation}
which has boundary conditions as follows
\begin{eqnarray}
 \lim_{r\rightarrow r_{H}}F(r)=(2n-1)m\pi,~~~ \lim_{r\rightarrow 0}F(r)=0, \label{bdy2}
\end{eqnarray}
for positive integer $n$, and $dr/d\mu >0$ implies that
$F^{-1}(\mu)$ is well-defined function.  We can thus rewrite the
GMGHS metric (\ref{inside}) with the lapse function
(\ref{schlapse2})
\begin{eqnarray}\label{nsmetric2}
ds^{2}&=&-d\mu^2+\Bigl({\frac{2m}{F^{-1}(\mu)}}-1\Bigr)dr^2
          +\Bigl({F^{-1}(\mu)}^2-\alpha{F^{-1}(\mu)}\Bigr)d\Omega^2\nonumber\\
      &=&-d\mu^2+f_1(\mu)^2dr^2+f_2(\mu)^2d\Omega^2
\end{eqnarray}
by using the warping functions (\ref{schf1f2}). Note that the
GMGHS metric in the multiply warped product spacetime has the same
form with the Kantowski-Sachs solution~\cite{KS} given by
\begin{equation}
ds^2=-dt^2+A^2(t)dr^2+B^2(t)\Omega^2,
\end{equation}
which represents homogeneous but anisotropically expanding
(contracting) cosmology.

Thus, in the case of the interior region $r<2m$, the GMGHS metric has
been rewritten as a multiply warped product spacetime having the
warping functions in terms of $f_1$ and $f_2$. Moreover, we can write
down the Ricci curvature on the multiply warped products as
\begin{eqnarray}
R_{\mu\mu}&=&-\frac{f''_1}{f_1}-\frac{2f''_2}{f_2},\nonumber\\
R_{tt}&=&f_1 f''_1+\frac{2f_1f'_1f'_2}{f_2},\nonumber\\
R_{\theta\theta}&=&\frac{f'_1f_2f'_2}{f_1}+f'^2_2+f_2f''_2+1,\nonumber\\
R_{\phi\phi}&=&\left(\frac{f'_1f_2f'_2}{f_1}+f'^2_2+f_2f''_2+1\right)\sin^2\theta,\nonumber\\
R_{mn}&=&0,~{\rm for}~m\neq n,
\end{eqnarray}
which have the same form with the Ricci curvature of the multiply
warped interior Schwarzschild metric~\cite{choi00}. The only
difference from the Schwarzschild is the $\alpha$ term in the
warping function $f_2$ in Eq.~{(\ref{schf1f2})}.


\section{Geodesic motion near hypersurface}
\setcounter{equation}{0}
\renewcommand{\theequation}{\arabic{section}.\arabic{equation}}

A full understanding of the GMGHS spacetime having an
event horizon with an essential singularity at the center and a
surface singularity at $r=\alpha$, etc, was recently achieved only
comparatively. Also, since the geodesics in the GMGHS spacetime
illuminate some basic aspects of an universe within the event
horizon, we shall include an account of them. In this section, we
briefly revisit the GMGHS interior spacetime with two warping
functions at a singular point $r=\alpha$ in the hypersurfaces, and
we are interested in investigating the geodesic curves of a static
spherically symmetric GMGHS spacetime near hypersurfaces.
\vskip10pt

In local coordinates $\{x^i\}$  the line element corresponding to
this metric (\ref{inside}) will be denoted by
\begin{eqnarray}
dS^2 = g_{ij}dx^idx^j. \label{rf7}
\end{eqnarray}
Consider the equations of geodesics in the GMGHS spacetime with
affine parameter $\lambda$ given by
\begin{eqnarray}
\frac{dx^i}{d\lambda^2}+
\Gamma^i_{jk}\frac{dx^j}{d\lambda}\frac{dx^k}{d\lambda} = 0.
\label{rf8}
\end{eqnarray}

Let a geodesic $\gamma$ be given by $\gamma(\tau) =
\Bigl(\mu(\tau), r(\tau), \theta(\tau), \phi(\tau)\Bigr)$ of the
interior GMGHS spacetime in the case of $r<2m$ from
Eq.~(\ref{inside}), then the orbits of the geodesics equation are
given as follows

\begin{eqnarray} \frac{d^2\mu}{d\tau^2}+f_1{\frac{df_1}{d\mu}}\left(\frac{dr}{d\tau}\right)^2+f_2\frac{df_2}{d\mu}\left(\frac{d\theta}{d\tau}\right)^2+f_2\frac{df_2}{d\mu}\sin^2\theta\left(\frac{d\phi}{d\tau}\right)^2&=&0,\\
\frac{d^2r}{d\tau^2}+\frac{2}{f_1}\frac{df_1}{d\tau}\frac{dr}{d\tau}&=&0,\\
\frac{d^2\theta}{d\tau^2}+\frac{2}{f_2}\frac{df_2}{d\tau}\frac{d\theta}{d\tau}-\sin\theta\cos\theta\left(\frac{d\phi}{d\tau}\right)^2&=&0,\\
\frac{d^2\phi}{d\tau^2}+\frac{2}{f_2}\frac{df_2}{d\tau}\frac{d\phi}{d\tau}+{2\cot\theta}\frac{d\theta}{d\tau}\frac{d\phi}{d\tau}&=&0
\end{eqnarray}
with a following constraint along the geodesic
\begin{equation}
-\left(\frac{d\mu}{d\tau}\right)^2+f^2_1\left(\frac{dr}{d\tau}\right)^2+f^2_2\left(\frac{d\theta}{d\tau}\right)^2+f^2_2\sin^2\theta
\left(\frac{d\phi}{d\tau}\right)^2=\varepsilon.
\end{equation}
Note that a timelike (nulllike) geodesic is taken as
$\varepsilon=-1~(\varepsilon=0)$.

Hereafter, without loss of generality, suppose the geodesic
\begin{equation}
\gamma(\tau_0 ) = \Bigl(\mu(\tau_0 ), r(\tau_0 ), \theta(\tau_0),
\phi(\tau_0)\Bigr)
\end{equation}
for some $\tau_0$ and the equatorial plane of
$\theta=\frac{\pi}{2}$, thus $\frac{d\theta}{d\tau}=0$. Then, the
geodesic equations are reduced to
\begin{eqnarray} \label{mueq}\frac{d^2\mu}{d\tau^2}+f_1\frac{df_1}{d\mu}\left(\frac{dr}{d\tau}\right)^2+f_2\frac{df_2}{d\mu}\left(\frac{d\phi}{d\tau}\right)^2&=&0,\\
\label{req}\frac{d^2r}{d\tau^2}+\frac{2}{f_1}\frac{df_1}{d\tau}\frac{dr}{d\tau}&=&0,\\
\frac{d^2\theta}{d\tau^2}&=&0,\\
\label{phieq}\frac{d^2\phi}{d\tau^2}+\frac{2}{f_2}\frac{df_2}{d\tau}\frac{d\phi}{d\tau}&=&0
\end{eqnarray}
with a constraint
\begin{equation}\label{constrainteq}
-\left(\frac{d\mu}{d\tau}\right)^2+f^2_1\left(\frac{dr}{d\tau}\right)^2+f^2_2\left(\frac{d\phi}{d\tau}\right)^2=\varepsilon.
\end{equation}
These geodesic equations can be simplified as follows
\begin{eqnarray}
\label{req1}\frac{dr}{d\tau}&=&\frac{c_1}{f_1^2},\\
\label{phieq1}\frac{d\phi}{d\tau}&=&\frac{c_2}{f_2^2},\\
\frac{d^2\theta}{d\tau^2}&=&0\\
\end{eqnarray}
with a constraint
\begin{eqnarray}
\label{constrainteq1}-\left(\frac{d\mu}{d\tau}\right)^2+\frac{c_1^2}{f_1^2}+\frac{c_2^2}{f_2^2}&=&\varepsilon.
\end{eqnarray}
The constant $c_1$ represents the total energy per unit rest mass
of a particle as measured by a static observer~\cite{gad,Clarke,Wald}, and $c_2$
represents the angular momentum in the GMGHS spacetimes. The
equations for $r$ and $\phi$ are obtained from Eqs.~(\ref{req}) and
(\ref{phieq}), respectively. Making use of these $r$, $\phi$
equations, we can show that Eq.~(\ref{mueq}) is the exactly same
with Eq.~(\ref{constrainteq}) when we take the integration
constant as $-\varepsilon/2$. \vskip5pt


Now, we recall the GMGHS spacetime $M=(a, b)\times_{f_1}R\times_{f_2}
S^2$ in the framework of the Lorentzian multiply warped products.
Let subspace $\Sigma_{x^i}$ of the GMGHS spacetime $M$ be a
regularly embedded $x^i$-directed hypersurface having coordinate
neighborhood $U(p)$ with local coordinates $(x^1, x^2, x^3, x^4)$ such
that $\Sigma_{x^i}\cap U = \{(x^1, x^2, x^3, x^4)\in U\mid
x^i=p\}$ for all $p\in \Sigma_{x^i}$. For convenience, we say that
such a neighborhood $U$ is partitioned by $\Sigma_{x^i}$.
\vskip10pt

First of all, we consider the null geodesics in the $r$-direction,
which is defined by the hypersurface $\Sigma_{r}$ by taking
$d\theta=d\phi=0$. Then, we have $c_2=0$ in Eq.~(\ref{phieq1}).
Two equations (\ref{req1}) and (\ref{constrainteq1}) are now reduced
to give
\begin{equation}\label{rgeod}
dr=\frac{d\mu}{f_1(\mu)},
\end{equation}
which can be solved to
\begin{equation}
 \mu=\sqrt{(2m-r)r}+m\tan^{-1}\left(\frac{r-m}{\sqrt{(2m-r)r}}\right)+\frac{m\pi}{2}\equiv h(r).
\end{equation}
We draw the null geodesics in the $r$-direction in Fig. 1, with the
mass parameters of $m=1$ and $m=2$. In this Figure, we see the
radial coordinate $r=h^{-1}(\mu)$ is a monotonic function of
$\mu$. Therefore, as the time flows between $0<\mu <m\pi$, the
radial coordinate increases monotonically as $0<r<2m$. \vskip10pt
\begin{figure}[t!]
   \centering
   \epsfbox{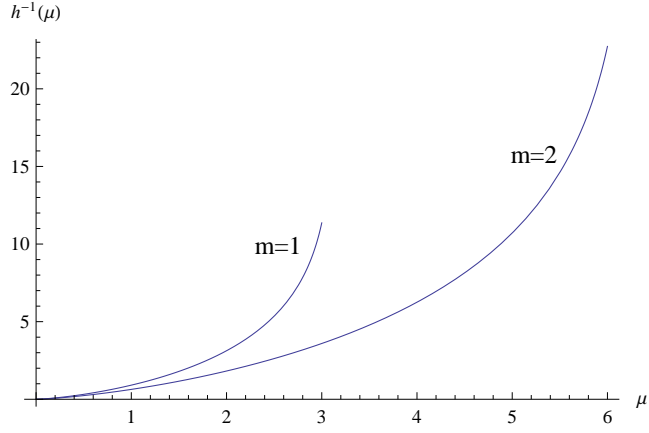}
\caption{The null geodesic $r(\mu)=h^{-1}(\mu)$ with the mass
parameters, $m=1,~2$, on the $\Sigma_{r}$ hypersurface.}
\label{fig1}
\end{figure}

Next, let us consider the geodesic in the $\phi$-direction, which
lies on the hypersurface $\Sigma_{\phi}$ at $\theta=\frac{\pi}{2}$
with $dr=0$.  Then, we have $c_1=0$ in Eq.~(\ref{req1}). Two
equations (\ref{phieq1}) and (\ref{constrainteq1}) are reduced to
give
\begin{equation}\label{phiDeq}
d\phi=\frac{d\mu}{f_2(\mu)},
\end{equation}
where $f_2(\mu)$ is given by Eq.~(\ref{fs}). In Fig. 2, we have
numerically drawn the azimuth angle. The left panel is drawn for
$m=1,~2$ with $\alpha=0$, which corresponds to the zero charged
pure Schwarzschild limit. On the other hand, the right panel is
drawn for $\alpha=0.5,~1$ with $m=1$. Here we note that the
azimuth angles start to appear at $\alpha=0.5,~1$, respectively,
for the first time. \vskip10pt
\begin{figure}[t!]
   \centering
   \epsfbox{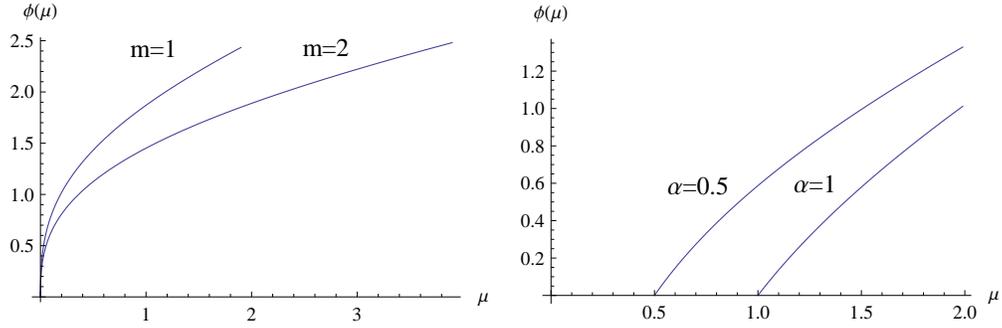}
\caption{Geodesic curve $\phi(\mu)$ at the plane
$\theta=\frac{\pi}{2}$ with $dr=0$: Left panel is for $m=1,~2$ with
$\alpha=0$, while right panel is  for $\alpha=0.5,~1$ with $m=1$.}
\label{fig2}
\end{figure}

Finally, let us find the geodesic in the $\mu$-direction, which is
defined by the hypersurface $\Sigma_{\mu}$, eliminating $\mu$ in
Eqs. (\ref{rgeod}) and (\ref{phiDeq}), leading to
\begin{equation}
d\phi=\frac{1}{r}\sqrt{\frac{2m-r}{r-\alpha}}dr.
\end{equation}
This has a solution as
\begin{equation}
\phi(r)=\sqrt{\frac{2m}{\alpha}}\cot^{-1}\left(\frac{2\sqrt{2m\alpha(2m-r)(r-\alpha)}}{2mr-4m\alpha+r\alpha}\right)
      +\tan^{-1}\left(\frac{2(m-r)+\alpha}{2\sqrt{(2m-r)(r-\alpha)}}\right).
\end{equation}
In Fig. 3, we have drawn the geodesic curve $\phi(r)$ for
$\alpha=0.5,~1$ on the hypersurface $\Sigma_{\mu}$.
\begin{figure}[t!]
   \centering
   \epsfbox{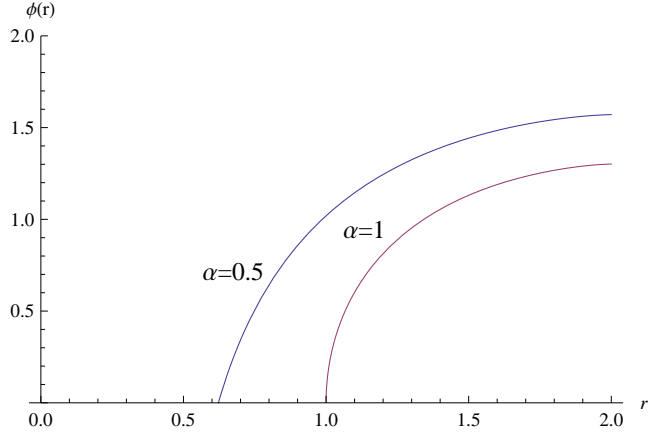}
\caption{Geodesic curves $\phi(r)$ for $m=1$ with $\alpha=0.5,~1$ on
the hypersurface $\Sigma_{\mu}$.} \label{fig3}
\end{figure}

\section{Conclusions}
\setcounter{equation}{0}
\renewcommand{\theequation}{\arabic{section}.\arabic{equation}}
\label{sec:conclusions}
In this paper, we have studied the GMGHS interior spacetime in
associated with a multiply warped product manifold. In the
multiply warped product manifold, the GMGHS spacetime has been
characterized by two warping functions $f_1(\mu)$ and $f_2(\mu)$,
compared with the Schwarzschild spacetime which has the one
warping function of $f_1(\mu)$.

We have also investigated the geodesic motions in the multiply
warped product spacetime near the hypersurfaces for each
directions. As results, in the multiply warped product spacetime,
we have shown the $r$-directed geodesic motion $r=h^{-1}(\mu)$
monotonically increases as $\mu$, while $\phi$-directed geodesic
motion has no azimuth angle when $r=F^{-1}(\mu)$ is smaller than
$\alpha$ and starts to appear when $r\ge\alpha$. We have also
obtained the most general geodesic curve $\phi(r)$ with $\alpha$
along the $r$ variation.

\section*{Acknowledgement}

J. Choi would like to acknowledge financial support from Korea Air
Force Academy Grant (KAFA 13-04).


\begin{thebibliography}{99}
\bibitem{bo} R.L. Bishop and B. O'Neill, Trans. A.M.S. {\bf 145}, 1 (1969).
\bibitem{be} J.K. Beem, P. E. Ehrlich and K. Easley,
            {\it Global Lorentzian Geometry} (Marcel Dekker Pure and Applied Mathematics, New York, 1996).
\bibitem{bpe} J.K. Beem and P. E. Ehrlich, Math. Proc. Camb. Phil.
          Soc. {\bf 85}, 161 (1979).
\bibitem{choi00} J. Choi, J. Math. Phys. {\bf 41}, 8163 (2000).
\bibitem{choi01} J. Choi and S.T. Hong, J. Math. Phys. {\bf 45}, 642 (2004).
\bibitem{choi02} S.T. Hong, J. Choi and Y.J. Park,  Nonlinear Analysis {\bf 63}, e493 (2005).
\bibitem{choi03} S.T. Hong, J. Choi and Y.J. Park,  Gen. Rel. Grav. {\bf 35}, 2105 (2003).
\bibitem{Horowitz1992}
     G.~T.~Horowitz, In Trieste 1992, Proceedings, String theory and quantum gravity '92, 55-99, hep-th/9210119.
\bibitem{gm} G.W. Gibbons and K. Maeda, Nucl. Phys. B {\bf 298}, 741 (1988).
\bibitem{ghs} D. Garfinkle, G.T. Horowitz and A. Strominger, Phys. Rev. D {\bf 43}, 3140 (1991).
\bibitem{hm08} K.~Hioki and U.~Miyamoto, Phys.\ Rev.\ D {\bf 78}, 044007 (2008).
\bibitem{gad} R. M, Gad,  Astrophys. Space Sci, {\bf 330}, 107 (2010).
\bibitem{Fernando}  S.~Fernando,
  Phys.\ Rev.\ D {\bf 85}, 024033 (2012).
\bibitem{GSW} M. B. Green, J. H. Schwarz, and E. Witten,
  {\it Supersting Theory}, (Cambridge University Press, Cambridge,  1987).
\bibitem{Nodvik}
  J.~S.~Nodvik,
  Phys.\ Rev.\ Lett.\  {\bf 55}, 2519 (1985).  
\bibitem{liu1} H. Liu and P. S. Wesson, Phys. Lett. {\bf B381}, 420 (1996).
\bibitem{liu2} H. Liu and P. S. Wesson, Class. Quantum Grav. {\bf 14}, 1651 (1997).
\bibitem{KS} R. Kantowski and R.K. Sachs, J. Math. Phys. {\bf 7}, 443 (1966).
\bibitem{Clarke} C. J. S. Clark, ¡±Elementary General relativity¡± (Edward Arnold,
London, 1979).
\bibitem{Wald} R. M. Wald, ¡±General Relativity¡± (University of Chicago Press, Chicago, 1984).
\end{thebibliography}
\end{document}